
%
%
%
\def\@{{\char'100}}

\long\def\abstract#1{\bigskip{\advance\leftskip by 2true cm
\advance\rightskip by 2true cm\eightpoint\centerline{\bf
Abstract}\everymath{\scriptstyle}\vskip10pt\vbox{#1}}\bigskip}
\long\def\resume#1{{\advance\leftskip by 2true cm
\advance\rightskip by 2true cm\eightpoint\centerline{\bf
R\'esum\'e}\everymath{\scriptstyle}\vskip10pt \vbox{#1}}}

\def\references{\bigbreak\centerline{\sc
References}\medskip\nobreak\bgroup
\def\ref##1&{\leavevmode\hangindent45pt
\hbox to 42pt{\hss\bf[##1]\ }\ignorespaces}
\parindent=0pt
\everypar={\ref}\par}
\def\endreferences{\egroup}
\long\def\authoraddr#1{\medskip{\baselineskip9pt\let\\=\cr
\halign{\line{\hfil{\Addressfont##}\hfil}\crcr#1\crcr}}}
\def\Subtitle#1{\medbreak\noindent{\Subtitlefont#1.} }
%
%
\newif\ifrunningheads
\runningheadstrue
\immediate\write16{- Page headers}
\headline={\ifrunningheads\ifnum\pageno=1\hfil\else\ifodd\pageno\rightheadline
\else\leftheadline\fi\fi\else\hfil\fi}
\def\rightheadline{\sc\hfil\RightHeadText\hfil}
\def\leftheadline{\sc\hfil\LeftHeadText\hfil}

\hyphenation{Darboux Riemannian}
%
%
\immediate\write16{- Fonts "Small Caps" and "EulerFraktur"}
%
%
%

\let\sc=\tensmc
%
%
\font\teneuf=eufm10  \font\seveneuf=eufm7 \font\fiveeuf=eufm5
\newfam\euffam \def\gr{\fam\euffam\teneuf}
\textfont\euffam=\teneuf \scriptfont\euffam=\seveneuf
\scriptscriptfont\euffam=\fiveeuf
%

\def \wt {\widetilde}

\def \ra {\rightarrow}

\def \lra {\longrightarrow}
\def \lmt {\longmapsto}
\def \a {\alpha}

\def \l {\lambda}

\def \m  {\mu}

\def \th {\theta}

\def \o {\omega}

\def \z {\zeta}

\def \ss {\subset}

\def \di {\partial}

\def \tr{{\rm tr}}
\def \det{{\rm det}}
\def\nchi{\hbox{\raise 2.5pt\hbox{$\chi$}}}
\hyphenation{Harnad}
%
%

\def\nchi{\hbox{\raise 2.5pt\hbox{$\chi$}}}
%
%

\def\CC{{\cal C}}
\def\DD{{\cal D}}
\def\EE{{\cal E}}

\def\II{{\cal I}}
\def\JJ{{\cal J}}

\def\NN{{\cal N}}
\def\OO{{\cal O}}
\def\PP{{\cal P}}

%
%

		\def\bfR{{\bf R}}

%
%
\def\authorfont{\sc}
\font\eightrm=cmr8
\font\eightbf=cmbx8
\font\eightit=cmti8
\font\eightsl=cmsl8

\def\eightpoint{\let\rm=\eightrm \let\bf=\eightbf \let\it=\eightit
\let\sl=\eightsl \baselineskip = 9.5pt minus .75pt  \rm}

\font\titlefont=cmbx10 scaled\magstep2
\font\sectionfont=cmbx10
\font\Subtitlefont=cmbxsl10
\font\Addressfont=cmsl8
%
%
\def\Satz#1:#2\par{\smallbreak\noindent{\sc #1:\ }
{\sl #2}\par\smallbreak}
%
%
\immediate\write16{- Section headings}
\newcount\secount
\secount=0
\newcount\eqcount
\outer\def\section#1.#2\par{\global\eqcount=0\bigbreak
\ifcat#10
 \secount=#1\noindent{\sectionfont#1. #2}
\else
 \advance\secount by 1\noindent{\sectionfont\number\secount. #2}
\fi\par\nobreak\medskip}
%
%
\immediate\write16{- Automatic numbering}
\catcode`\@=11
\def\adv@nce{\global\advance\eqcount by 1}
\def\unadv@nce{\global\advance\eqcount by -1}
\def\nextnumber{\adv@nce}
%
%
\newif\iflines
\newif\ifm@resection
\def\onesec{\m@resectionfalse}
\def\moresec{\m@resectiontrue}
\moresec
\def\eq{\global\linesfalse\eq@}
\def\eqn{\global\linestrue&\eq@}
\def\nosubind@x{\global\subind@xfalse}
\def\newsubind@x{\ifsubind@x\unadv@nce\else\global\subind@xtrue\fi}
\newif\ifsubind@x
\def\eq@#1.#2.{\adv@nce
 \if\relax#2\relax
  \edef\loc@lnumber{\ifm@resection\number\secount.\fi
  \number\eqcount}
  \nosubind@x
 \else
  \newsubind@x
  \edef\loc@lnumber{\ifm@resection\number\secount.\fi
  \number\eqcount#2}
 \fi
 \if\relax#1\relax
 \else
  \expandafter\xdef\csname #1@\endcsname{{\rm(\loc@lnumber)}}
  \expandafter
  \gdef\csname #1\endcsname##1{\csname #1@\endcsname
  \ifcat##1a\relax\space
  \else
   \ifcat\noexpand##1\noexpand\relax\space
   \else
    \ifx##1$\space
    \else
     \if##1(\space
     \fi
    \fi
   \fi
  \fi##1}\relax
 \fi
 \eq@@{\loc@lnumber}}
\def\eq@@#1{\iflines \else \eqno\fi{\rm(#1)}}
\def\m@th{\mathsurround=0pt}
%
%
\def\display#1{\null\,\vcenter{\openup1\jot
\m@th
\ialign{\strut\hfil$\displaystyle{##}$\hfil\crcr#1\crcr}}
\,}
\newif\ifdt@p
\def\@lign{\tabskip=0pt\everycr={}}
\def\displ@y{\global\dt@ptrue \openup1 \jot \m@th
 \everycr{\noalign{\ifdt@p \global\dt@pfalse
  \vskip-\lineskiplimit \vskip\normallineskiplimit
  \else \penalty\interdisplaylinepenalty \fi}}}
%
%
\def\displayno#1{\displ@y \tabskip=\centering
 \halign to\displaywidth{\hfil$
\@lign\displaystyle{##}$\hfil\tabskip=\centering&
\hfil{$\@lign##$}\tabskip=0pt\crcr#1\crcr}}
%
%
\def\cite#1{{\bf[#1]}}
\catcode`\@=\active
%
%
\magnification=\magstep1
\hsize= 6.75 true in
\vsize= 8.75 true in
%
%
\def\RightHeadText{Classical and Quantum Separability}
\def\LeftHeadText{J. Harnad and P. Winternitz}
%
%
\rightline{CRM-1921 (1993) \break}
\bigskip
\centerline{\titlefont Classical and Quantum Integrable Systems in}
\centerline{\titlefont $\wt{\gr{gl}}(2)^{+*}$ and Separation of
Variables\footnote{${}^{\dagger}$}{\eightpoint
Research supported in part by the Natural Sciences and
Engineering Research Council of Canada and the Fonds FCAR du Qu\'ebec.}}
\bigskip
\centerline{\authorfont J.~Harnad\footnote{$^{\scriptstyle1}$}{\eightpoint
e-mail address: harnad\@alcor.concordia.ca {\it\ or\ }
harnad\@mathcn.umontreal.ca}}
\authoraddr
{Department of Mathematics and Statistics, Concordia University\\
7141, Sherbrooke W., Montr\'eal, Qu\'ebec, Canada H4B 1R6, and\\
Centre de recherches math\'ematiques, Universit\'e de Montr\'eal\\
C.~P.~6128, succ.``A'', Montr\'eal, Qu\'ebec, Canada H3C 3J7}
\bigskip
\centerline{\authorfont  P.~Winternitz\footnote{$^{\scriptstyle2}$}
{\eightpoint e-mail address: wintern\@ere.umontreal.ca}}
\authoraddr
{Centre de recherches math\'ematiques, Universit\'e de Montr\'eal\\
C.~P.~6128, succ.``A'', Montr\'eal, Qu\'ebec, Canada H3C 3J7}
\bigskip
\abstract{Classical integrable Hamiltonian systems generated by elements of
the Poisson commuting ring  of spectral invariants on rational coadjoint
orbits of the loop algebra $\wt{\gr{gl}}^{+*}(2,{\bf R})$ are integrated by
separation of variables in the Hamilton-Jacobi equation in hyperellipsoidal
coordinates.  The canonically quantized systems are then shown to also be
completely integrable and separable within the same coordinates. Pairs of
second
class constraints defining reduced phase spaces are implemented in the
quantized systems by choosing one constraint as an invariant, and interpreting
the other as determining a quotient (i.e., by treating one as a first class
constraint and the other as a gauge condition). Completely integrable,
separable
systems on spheres and ellipsoids result, but those on ellipsoids require a
further modification of order $\OO(\hbar^2)$ in the commuting invariants in
order
to assure self-adjointness and to recover the Laplacian for the case of free
motion. For each case - in the ambient space ${\bf R}^{n}$, the sphere and the
ellipsoid - the Schr\"odinger equations are completely separated in
hyperellipsoidal coordinates, giving equations of generalized Lam\'e type. }
\bigskip
\baselineskip 14 pt
\noindent{\sectionfont Introduction}\medskip\nobreak

   A general method for realizing integrable Hamiltonian systems as isospectral
flows in rational coadjoint orbits of loop algebras was developed in \cite{AHP,
AHH1-AHH4}. This approach begins with a moment map embedding of certain
Hamiltonian quotients of symplectic vector spaces into finite dimensional
Poisson
subspaces of the dual $\wt{\gr gl}(r)^{+*}$ of the positive frequency part of
the
loop algebra $\wt{\gr gl}(r)$  (or certain subalgebras thereof). The
Adler-Kostant-Symes (AKS) theorem \cite{A, K, S} then implies that the spectral
invariants provide commuting integrals inducing isospectral flows
determined by matrix Lax equations. The level sets of these commuting
invariants
are shown to determine Lagrangian foliations on the rational coadjoint orbits,
and hence completely integrable systems. Finally, a special set of canonical
coordinates, the {\it spectral Darboux coordinates} are introduced, in which
the
Liouville generating function, which determines the linearizing canonical
transformation, is expressed in completely separated form as an abelian
integral
on the associated invariant spectral curve. The resulting linearizing map is
essentially the Abel map to the Jacobi variety of the spectral curve, thus
providing a link, through purely Hamiltonian methods, with the
algebro-geometric
linearization methods of \cite{Du, KN, AvM}. This approach has been applied to
the study of a large number of integrable classical Hamiltonian systems, as
well
as the determination of finite dimensional quasi-periodic solutions of
integrable
systems of PDE's \cite{H, HW, AHH3, AHH4, W, TW}.

   In the present work we focus on the case $\wt{\gr gl}(2)^{+*}$,
taking an equivalent approach to integrability based, first of all, on
separation of variables in the Hamilton-Jacobi equation. The relevent
``spectral
Darboux coordinates'' in this case simply reduce to hyperellipsoidal
coordinates.
The purpose of this reformulation is to prepare the passage to the
corresponding
quantum systems and the study of integrability and separation of variables in
the
associated Schr\"odinger equation. As it turns out, each such classical
integrable system has an integrable quantum analogue, for which the
Schr\"odinger equation is completely separable in the same coordinates. One
case
of separation of variables in such systems, - the quantized Neumann oscillator
(an anisotropic harmonic oscillator constrained to the surface of a
sphere) - was studied in \cite{BT}, and the results extended to the quantum
Rosochatius system in \cite{Mc}. Other special cases, involving quantized free
motion in various symmetric spaces and reductions thereof, were studied in
\cite{K, KM, KMW, BKW1, BKW2, ORW, Ku, To}. All these systems may be placed in
a
loop algebra setting using the moment map embedding of \cite{AHP}, and
canonically quantized. The resulting formulation is equivalent to a Gaudin spin
chain \cite{G, Ku}, with $\gr{su}(2)$ replaced by $\gr{gl(2)}$, and the
separation
of variables interpreted as a ``functional Bethe ansatz'' \cite{Sk1, Sk2}.

In \cite{Mo}, the algebraic geometry of a number of classical integrable
systems constrained to quadrics in ${\bf R}^n$ was examined. The integration of
these and related systems was given a loop  algebra formulation based on
$\wt{\gr
gl}^+(2,{\bf C})$ and reductions thereof in \cite{AHP, AHH4}. In the
present work, such systems will be reexamined in terms of separation of
variables
in the Hamilton-Jacobi equation. Their quantum analogues will then be studied
through constrained canonical quantization, making use of the loop algebra
formulation to identify the commuting invariants in terms of ``quantum
determinants''. The corresponding Schr\"odinger equations will be shown to
separate within the same coordinates as the classical systems. Constraints
leading to dynamics on spheres and ellipsoids will be  shown to lead to
integrable quantum systems, also separable in the same coordinates as the
classical ones.

  In Section 1, the appropriate loop algebra formulation of the systems in
question is given. In each case, the coadjoint orbit is identified with the
quotient $\bf R^{2n}/({\bf Z}_2)^n$, and integrable isospectral flows are
examined both in this space and on constrained submanifolds identified with
the cotangent bundle of a sphere $S^{n-1}\ss \bfR^n$, or an ellipsoid
$\EE^{n-1}\ss {\bf R}^n$. The key step consists of using a Lagrange
interpolation formula to express the invariant spectral polynomial in terms
of its values at the associated spectral divisor points, and
noting that these values coincide with the squared canonical momentum
components. The separation of variables follows from identification of residues
in the interpolation formula.

   In Section 2, the corresponding quantum systems are obtained by canonical
quantization in the ambient phase space ${\bf R}^{2n}$ before quotienting. The
resulting Schr\"odinger equation is again expressed in terms of
hyperellipsoidal coordinates through Lagrange interpolation, and the
completely separated form is deduced, again by identification of residues,
giving various types of generalized Lam\'e equations. The associated one
dimensional Schr\"odinger operators are seen to give the quantized form of the
invariant spectral curves (cf. \cite{Sk3}). In the case of the constrained
systems
on the sphere $S^{n-1}$ the same formulation, together with second class
constraints, leads without difficulty to completely separable integrable
systems. In the case of the ellipsoid ${\cal E}^{n-1}$, however, a new problem
arises, since the separation of variables, while  holding for the Schr\"odinger
equation, does not hold for the volume element, giving rise at first to non
self-adjoint operators. This is easily rectified by noting that the resulting
Schr\"odinger operators are nevertheless self-adjoint with respect to a scalar
product determined by a different measure than the one associated to the
induced
volume form. Conjugating the operators by the map relating the two scalar
products, self-adjoint operators are obtained with respect to the standard
measure. However, to recover the Laplacian in the case of free motion, a
further
scalar term  of order $\OO(\hbar^2)$ must be added to the quantum Hamiltonians.
Since this correction gives the same semi-classical limit and does not destroy
the
integrability (or separability), it may be viewed as a satisfactory quantized
version of the associated classical integrable systems.
\bigskip
\section 1. Classical Systems in $\wt{\gr gl}(2)^{+*}$
\medskip
\Subtitle {1a. Ambient Space}

Following the general approach of \cite{AHP, AHH2, H}, we define a Poisson map
$$
\eqalignno{
\wt{J}_A&:{\bf R}^{2n}:\lra \wt{\gr gl}(2)^{+*}  \eqn JAa.a.\cr
\wt{J}_A&:({\bf x, y})\lmt \NN_0(\l) \eqn JAb.b.}
$$
where
$$
\NN_0(\l):= {1\over 2} \pmatrix{-\sum_{i=1}^n {x_i y_i -\m_i \over \l-\a_i}&
     \sum_{i=1}^n {y_i^2 -{\m_i\over x_i^2} \over \l -\a_i}\cr
     \sum_{i=1}^n {x_i^2 \over \l -\a_i} &
   \sum_{i=1}^n {x_i y_i + \m_i \over \l -\a_i}}, \eq NM..
$$
${\bf x, y}\in {\bf R}^n$ have components $(x_i,y_i)_{i=1,\dots n}$ and
$\{\mu_i, \a_i\}_{i=1, \dots n}$ is a set of $2n$ arbitrary real constants,
the $\a_i$'s being chosen as distinct. Here, the loop algebra $\wt{\gr gl}(2)$
consists of smooth maps $X:S^1\lra {\gr gl}(2)$ from a fixed circle $S^1$,
centred at the origin of the complex $\l$--plane, to  ${\gr gl}(2)$, the
subalgebra $\wt{\gr gl}(2)^+$ consists of elements $X\in \wt{\gr gl}(2)$
admitting a holomorphic extension to the interior of $S^1$, and the (smooth)
dual space  $\wt{\gr gl}(2)^{+*}$ is identified with the subalgebra
$\wt{\gr gl}(2)^-$ of elements $X$ admitting a holomorphic extension outside
$S^1$, with  $X(\infty)=0$. The dual pairing $<\ , \ >$ is defined by
integration:
$$
\eqalignno{
<X,Y>&:= {1\over 2\pi i} \oint_{S^1}\tr \left(X(\l) Y(\l)\right) d\l \eqn..\cr
X\in \wt{\gr gl}^{+*}(2) &\sim \wt{\gr gl}(2)^-,
\quad Y \in \wt{\gr gl}^{+}(2). }
$$
To assure that $\NN_0 \in \wt{\gr gl}(2)^{+*}$, we must choose the constants
$\{\a_i\}$ entering in the definition \NM of the Poisson map $\wt{J}_A$ to
be in the interior of $S^1$. The image of the map is a coadjoint orbit in
$\wt{\gr gl}(2)^{+*}$, identified with the quotient
${\bf R}^{2n}/({\bf Z}_2)^n$ of the phase space ${\bf R}^{2n}$ by the
symplectic action of the group $({\bf Z})_2^n$  of reflections in the
coordinate hyperplanes, and $\{x_i,y_i\}_{i=1,\dots n}$ are canonical
coordinates defining the standard symplectic form
$$
\o = \sum_{i=1}^n dx_i\wedge dy_i. \eq..
$$

  Fixing an element
$$
Y=
\pmatrix{a & b \cr
          c & -a} \in sl(2, {\bf R}), \eq..
$$
we let
$$
\eqalignno{
\NN(\l) &:= Y + \NN_0(\l) \cr
  & = \pmatrix{a & b \cr
               c & -a}  +
{1\over 2} \pmatrix{-\sum_{i=1}^n {x_i y_i -\m_i \over \l-\a_i}&
     \sum_{i=1}^n {y_i^2 -{\m_i\over x_i^2} \over \l -\a_i}\cr
     \sum_{i=1}^n {x_i^2 \over \l -\a_i} &
   \sum_{i=1}^n {x_i y_i + \m_i \over \l -\a_i}}\in \wt{\gr gl}(2).
\eqn Nlambda..}
$$
The ring $\II^Y_{AKS}$ of commuting invariants is chosen, according to
the Adler-Kostant-Symes (AKS) theorem, by restricting the ring
$\II(\wt{\gr gl}^*(2))$ of $Ad^*$ invariants on the dual space
$\wt{\gr gl}^*(2)$ of the full loop algebra to the translate of the coadjoint
orbit  $\OO_{\NN_0} \ss \wt{\gr gl}(2)^{+*}\ss \wt{\gr gl}(2)^*$ by the
fixed element $Y$. Picking any element $\phi \in \II^Y_{AKS}$ as Hamiltonian,
the AKS theorem implies that the equations of motion take the Lax form
$$
{\di \NN \over \di t} =
[(d\phi)(\NN)_+, \NN] \eq..
$$
and hence determine an isospectral flow. Since $\II^Y_{AKS}$ is just the ring
of spectral invariants, fixing simultaneous level sets of its elements
amounts to fixing the coefficients of the characteristic polynomial
$$
\PP(\l,\z):=
\det(\NN(\l) -\z {\bf I}_2)  = \z^2 -\z \sum_{i=1}^n{\mu_i \over \l-\a_i} +
{P(\l) \over a(\l)}.  \eq Specpol..
$$
where
$$
\eqalignno{
\det(\NN(\l))
&:=  {P(\l) \over a(\l)} = {1\over 2}\sum_{i=1}^n {I_i\over \l -\a_i} -(a^2+b
c),   \eqn DetN.. \cr
a(\l) &:= \prod_{i=1}^n(\l -\a_i), \eqn Minpoly..}
$$
and
$$
I_i := {1\over 2} \sum_{j=1 \atop j\neq i}^n
{(x_i y_j - x_j y_i)^2 -\m_i^2{x_j^2 \over x_i^2}- \m_j^2{x_i^2 \over x_j^2}
+2\m_i \m_j \over \a_i -\a_j}
+2a x_i y_i - b x_i^2 + c\left(y_i^2 -{\m_i^2\over x_i^2}\right) \eq..
$$
are the (generalized) Devaney-Uhlenbeck invariants (cf. \cite{Mo} ).
The leading term of the polynomial
$$
P(\l) = \sum_{i=0}^n P_i \l^i \eq Psum..
$$
has constant coefficient
$$
P_n =-(a^2 + bc),  \eq Pn..
$$
while the remaining coefficients $\{P_0, \dots P_{n-1}\}$ are independent
generators of the ring of commuting invariants.

 Letting
$$
\eqalignno{
{1\over 2} \sum_{i=1}^n {\m_i \over \l -\a_i} &:= {K(\l) \over a(\l)}
\eqn.a.\cr
z &:= a(\l)(\z -{1\over 2} \sum_{i=1}^n{\m_i \over \l-\a_i}),
\eqn.b.}
$$
the invariant spectral curve $\CC$ is defined by
$$
z^2 = K^2(\l) - a(\l)P(\l), \eq Spectralcurve..
$$
and thus is hyperelliptic. Since $P(\l)$ has leading terms of the form
$$
P(\l) = -(a^2 +b c)\l^n + \left({1\over 2} \sum_{i=1}^n I_i+(a^2 +b
c)\sum_{i=1}^n \a_i\right) \l^{n-1} + \OO(\l^{n-2}), \eq Specpoly..
$$
$\CC$ generically has genus $g=n-1$ if  $a^2+bc \neq 0$ or
$\sum_{i=1}^n I_i \neq 0$ and $g=n-2$ if $a^2+b c =0$ and
$\sum_{i=1}^n I_i = 0$.

  In the following, instead of considering individual Hamiltonians within
the ring of spectral invariants, it will be convenient (as in \cite{BT}) to
treat the invariant polynomial $P(\l)$  as a one parameter linear family of
Hamiltonians, all commuting amongst themselves.  It will be
necessary to distinguish two cases, depending on whether $c=0$ or $c\neq 0$.
\medskip
\Subtitle {1b. Case (i) \ $c=0$. Restiction to $T^*S^{n-1}$}

Following the general procedure of  \cite{AHH4, H} the spectral Darboux
coordinates  \break $\{q,p, \l_{\mu}, \z_\mu\}_{\mu=1, \dots n-1}$ are defined
by
$$
\eqalignno{
\sum_{i=1}^n {x_i^2\over \l-\a_i} &= {e^q Q(\l) \over a(\l)} \eqn Darbia.a.\cr
Q(\l) &:=\prod_{\mu=1}^{n-1} (\l-\l_\mu), \eqn Darbib.b.\cr
e^q &:=\sum_{i=1}^n x_i^2. \eqn Darbic.c.}
$$
and
$$
\eqalignno{
\z_\mu &:= {1\over 2} \sum_{i=1}^n {x_i y_i \over \l_{\mu} - \a_i}.
\eqn Darbie.d.   \cr
p &:={1\over 2} \sum_{i=1}^n x_i y_i \eqn Darbid.e.}
$$
Here $\{q, \l_{\mu}\}_{\mu=1, \dots n-1}$ are, essentially, hyperellipsoidal
coordinates on ${\bf R}^n$, and $\{p,\z_\mu\}_{\mu=1, \dots n-1}$ are the
canonically conjugate momenta. To make such an identification, we
must also assume that the $\a_i$'s are all real and positive, and choose an
ordering such that, e.g.,
$$
\a_n <\l_{n-1}<\a_{n-1}<\l_{n-2} < \dots \l_1 <\a_1. \eq..
$$

   The canonical $1$--form on ${\bf R}^{2n}$ is
$$
\th = \sum_{i=1}^n y_idx_i = p dq + \sum_{\mu=1}^{n-1}\z_\mu d\l_\mu.
\eq Canformi..
$$
The Jacobian matrix of the coordinate change is defined by the partial
derivatives
$$
{\di x_i \over \di \l_{\mu}} = {1\over 2} {x_i \over \l_{\mu} - \a_i},
\quad {\di x_i \over \di q} = {x_i \over 2}, \eq Jacmatrixi..
$$
and its inverse by
$$
{\di \l_\mu \over \di x_i} = - {2 a(\l_\mu)\over Q'(\l_\mu)} {x_i\over \l_\mu
-\a_i}, \quad {\di q \over \di x_i} =2x_i e^{-q}. \eq InvJacmatrixi..
$$
The coordinate frame fields are thus
$$
\eqalignno{
{\di \over \di \l_\mu}& ={1\over 2} \sum_{i=1}^n {x_i\over \l_\mu
-\a_i}{\di\over \di x_i},\quad \mu=1, \dots n-1  \eqn Framefieldia.a. \cr
{\di \over \di q } &={1\over 2} \sum_{i=1}^n x_i{\di \over \di x_i}.
\eqn Framefieldib.b.}
$$

The ${\bf R}^n$ euclidean metric in these coordinates is
$$
\sum_{i=1}^ndx_i^2 = {1\over 4} e^q dq^2 -
{e^q\over 4} \sum_{\mu=1}^{n-1}{Q^\prime(\l_\mu)\over a(\l_\mu)} d\l_\mu^2
\eq Metrici..
$$
and the volume form is
$$
dV = {e^{nq\over 2}\over 2^n}{\prod_{\nu < \mu}^{n-1}(\l_\mu - \l_\nu)\over
   |\prod_{\mu} a(\l_\mu)|^{1\over 2}}\,dq \wedge d\l_1 \wedge \dots \wedge
d\l_{n-1}. \eq Volforma..
$$

  In order to express the linear family of Hamiltonians  generated by $P(\l)$
in terms of the hyperellipsoidal coordinates, we note first that it follows
from \Nlambda  and \Specpol that the values of $P(\l_\mu)$ are given by
$$
\eqalignno{
{P(\l_\mu) \over a(\l_\mu)} &= - (\z_\mu -a)^2 + {1\over 4}
\left(\sum_{i=1}^n{\mu_i \over \l_\mu - \a_i}\right)^2 \eqn Poveraa.a.\cr
& = - (\z_\mu -a)^2 + {K(\l_\mu)^2 \over a(\l_\mu)^2}. \eqn Poverab.b.}
$$
Using Lagrange interpolation, $P(\l)$ may then be expressed as
$$
\eqalignno{
P(\l)&=
 Q(\l)\left((-\l + \sum_{\mu=1}^{n-1} \l_\mu + \sum_{i=1}^n \a_i)a^2  +
{1\over 2} \sum_{i=1}^nI_i\right)\cr
&\qquad -\sum_{\mu=1}^{n-1}{Q(\l)a(\l_\mu)\over (\l -\l_\mu)Q'(\l_\mu)}
\left( (\z_\mu -a)^2-{K(\l_\mu)^2\over a(\l_\mu)^2}\right), \eqn Lagrinti..}
$$
where
$$
{1\over 2} \sum_{i=1}^n I_i = 2 a p -{b\over 2} e^q. \eq SumI..
$$

To write the Hamilton-Jacobi equation, we must reinterpret the coefficients
of the invariant polynomial $P(\l)$ in \Lagrinti, not as functions on the
phase space, but rather as integration constants and replace the
canonical momentum components $\{\z_\mu, p\}_{\mu=1, \dots n-1}$ by the
partial derivatives  $\{{\di S \over\di \l_\mu}, {\di S \over \di q}\}_{\mu=1,
\dots n-1}$ of the Hamilton characteristic function. The resulting form for
the time independent Hamilton-Jacobi equation is then
$$
\eqalignno{
&-\sum_{\mu=1}^{n-1}{Q(\l) a(\l_\mu)\over (\l -\l_\mu)Q'(\l_\mu)}
\left(\left({\di S \over \di \l_\mu} -a\right)^2
-{K(\l_\mu)^2\over a(\l_\mu)^2}\right) \cr
&\qquad + Q(\l)\left(-a^2\l + 2a{\di S \over \di q} -{b\over 2}e^q +
a^2(\sum_{i=1}^n \a_i - \sum_{\mu=1}^{n-1}\l_\mu)\right)
= P(\l),  \eqn HamJaci..}
$$
where the leading coefficient $P_n$ of $P(\l)$ is given in \Pn, and the
remaining $n$ coefficients $\{P_0,\dots, \dots,P_{n-1}\}$ are interpreted as
integration constants determining the ``energies'' for the parametric family of
Hamiltonians defined in \Lagrinti.

The integration of \HamJaci then proceeds by separation of variables.
Expressing $S$ in the separated form
$$
S(\l_1, \dots \l_{n-1}, q) = s_0(q)+ \sum_{\mu=1}^{n-1} s_\mu(\l_\mu), \eq
Slq..
$$
dividing both sides of \HamJaci by $Q(\l)$ and equating the leading terms in
$\l$ at $\infty$, as well as the residues at $\{\l=\l_\mu\}_{\mu=1, \dots
n-1}$ in the resulting equation gives
$$
\eqalignno{
\left({\di s_\mu \over \di \l_\mu} -a\right)^2 &= {K(\l_\mu)^2 - a(\l_\mu)
P(\l_\mu) \over a(\l_\mu)^2} \eqn Separia.a. \cr
2 a {\di s_0 \over \di q} &= {b\over 2} e^q - a^2 \sum_{i=1}^n\a_i + P_{n-1}.
\eqn Separib.b.}
$$
If $a\neq 0$, this may be integrated to give the completely separated solution
$$
S(\l_1, \dots, \l_{n-1}, q)= {b\over 4a} e^q + {q\over 2 a} (P_{n-1}
-a^2\sum_{i=1}^n a_i)+a\sum_{\mu=1}^{n-1} + \sum_{\mu=1}^n
\int_0^{\l_\mu}\sqrt{{K^2(\l)-a(\l)P(\l)\over a^2(\l)}}d\l.\eq..
$$
The linearizing coordinates are then
$$
\eqalignno{
Q_i &:= {\di S \over \di P_i} = {1\over 2}\sum_{\mu=1}^{n-1} \int_0^{\l_\mu}
{\l^i \over \sqrt{ K^2(\l) - a(\l)P(\l)}}d\l, \qquad i=0, \dots n-2
\eqn Abinta.a.\cr
Q_{n-1} &:= {\di S \over \di P_{n-1}} = {q\over 2a}
+{1\over 2}\sum_{\mu=1}^{n-1}\l_\mu \int_{0}^{\l_\mu} {\l^{n-1}\over \sqrt{
K^2(\l) - a(\l)P(\l)}}d\l.
 \eqn Abintb.b.}
 $$
The first $n-1$ of these, defined by eq. \Abinta, involve  abelian integrals of
the first kind and essentially define the Abel map to the Jacobi variety
$\JJ(\CC)$. The last one, defined by \Abintb, is an abelian integral of
the third kind, the integrand having a pair of simple poles over $\l=\infty$.
The linear flow induced by any Hamiltonian $\phi=\phi(P_0, \dots P_{n-1})$ in
the ring $\II^Y_{AKS}$ of spectral invariants is then given by
$$
Q_i = Q_{i0} + {\di \phi \over \di P_i}t, \qquad i=0, \dots n-1.  \eq Lin..
$$
By standard Jacobi inversion techniques (cf. \cite{Du, GH, AHH4}), any
function of the coordinates $\{\l_\mu, q\}_{\mu=1,\dots n-1}$ that is
symmetric in the $\l_\mu$'s, can be given an explicit form in terms of the
Riemann theta functions associated to the curve $\CC$.

  If $a=0$ and $b=0$, both $q$ and $p$ are conserved quantities, and
Hamilton's equations may be integrated on the invariant symplectic submanifold
given by fixing a level set of $q$ and $p$. By eqs. \Darbic, \Darbid, this
defines the cotangent bundle $T^*S^{n-1} \ss \bf{R}^{2n}$ to the sphere
$S^{n-1}\ss \bf{R}^n$. The Hamilton-Jacobi equation \HamJaci, may
then be interpreted on $S^{n-1}$, by choosing $S=S(\l_1, \dots, \l_{n-1})$.
The separated form is again given by \Slq, \Separia, with $s_0=0$ and
\Separib omitted. Both the leading and next to leading terms in eqs. \Lagrinti
vanish, so $P(\l)$ is of degree $n-2$. The completely separated solution is
$$
S(\l_1, \dots, \l_{n-1}, q)=  \sum_{\mu=1}^n
\int_0^{\l_\mu}\sqrt{{K^2(\l)-a(\l)P(\l)\over a^2(\l)}}d\l \eq..
$$
and the linearizing equations are given by \Abinta and \Lin, for $i=0,
\dots n-1$. Since the genus of $\CC$ is $g=n-2$, the $i=n-2$ integral in
\Abinta becomes singular, the integrand having simple poles over $\l=\infty$.

  If $a=0$ and $b\neq 0$, $q$ is still a conserved quantity, but $p$ is not.
Since the linear family of Hamiltonians $P(\l)$ now has no dependence on $p$,
to apply the Hamilton-Jacobi theory, the r\^oles of $p$ and $-q$ must be
interchanged, and the term $e^q$ in \HamJaci replaced by  $e^{\di S\over \di
p}$. The Hamilton characteristic function $S$ is now a function of
$\{\l_\mu, p\}_{\mu=1, \dots n-1}$ and the solution is obtained in
completely separated form as
$$
S(\l_1, \dots, \l_{n-1}, p) = p\,{\rm ln }\left({-2P_{n-1}\over b}\right) +
\sum_{\mu=1}^n \int_0^{\l_\mu}\sqrt{{K^2(\l)-a(\l)P(\l)\over a^2(\l)}}d\l.
\eq..
 $$
 The linear flow equations \Abinta and \Lin remain the same for $i=1, \dots
n-2$, while \Abintb is replaced by
$$
Q_{n-1} := {\di S \over \di P_{n-1}} = {p\over P_{n-1}}
+{1\over 2}\sum_{\mu=1}^{n-1} \int_{0}^{\l_\mu} {\l^{n-1}\over \sqrt{ K^2(\l) -
a(\l)P(\l)}}d\l, \qquad i=0, \dots n-2. \eq..
$$

   Even if $a\neq 0$, and $q$ and $p$ are not individually conserved
quantities, we may still impose the second class constraints
$$
\eqalignno{
\sum_{i=1}^n x_i^2 &= 1, \eqn Constra.a.\cr
\sum_{i=1}^n x_i y_i &=0,  \eqn Constrb.b.}
$$
\nextnumber
or, equivalently,
$$
\eqalignno{
q&=0,\eqn Constrq.a. \cr
 p&=0,\eqn Constrp.b.}
$$
which define the symplectic submanifold $T^*S^{n-1}\ss {\bf R}^{2n}$ as phase
space. The term $P_{n-1}$ in $P(\l)$ is no longer viewed as an
independent dynamic variable, or an integration constant, but rather the fixed
constant defined by
$$
P_{n-1} := a^2\sum_{i=1}^n\a_i -{b\over2}. \eq..
$$
The hyperellipsoidal coordinates $\{\l_\mu\}_{\mu=1,\dots n-1}$, given by
\Darbia, \Darbib are now interpreted as defined on $S^{n-1}\ss {\bf R}^n$
and, together with the conjugate momenta $\{\z_\mu\}_{\mu=1,\dots n-1}$,
these provide a canonical system on $T^*S^{n-1}$. The invariant coefficients
$\{P_0, \dots P_{n-2}\}$ of $P(\l)$ still form a Poisson
commutative set when constrained to $T^*S^{n-1}$, even though $q$ and $p$ are
not individually conserved quantities. (This follows from the fact that in the
ambient space, the combination  \SumI commutes with all the $P_i$'s.)

  For later use in Sec. 2, we note that the induced metric on $S^{n-1}$ in
the hyperellipsoidal coordinates $\{\l_1, \dots \l_{n-1}\}$ is
$$
\sum_{i=1}^ndx_i^2 \vert_{S^{n-1}}=
{1\over 4} \sum_{\mu=1}^{n-1}{Q^\prime(\l_\mu)\over a(\l_\mu)} d\l_\mu^2,
\eq MetricS..
$$
 the volume form on $S^{n-1}$ is
$$
dV_{S^{n-1}} = {1\over 2^n}{\prod_{\nu < \mu}^{n-1}(\l_\mu - \l_\nu)\over
   |\prod_{\mu} a(\l_\mu)|^{1\over 2}}\,d\l_1 \wedge \dots \wedge
d\l_{n-1}, \eq VolformS..
$$
and the scalar Laplacian is
$$
\Delta_{S^{n-1}}= 4\sum_{\mu=1}^{n-1}{a(\l_\mu)\over
Q'(\l_\mu)} \left({\di^2\over \di \l_\mu^2}
 + {1\over 2}\sum_{j=1}^n{1\over \l_\mu - \a_j}{\di\over \di\l_\mu} \right).
\eq LaplacianS..
$$

  The Hamilton-Jacobi equation is the same as in the unconstrained
case \HamJaci, but with the ${\di S \over \di q}$ term omitted. The solution
in completely separated form is again just
$$
S( \l_1, \dots, \l_{n-1})=  \sum_{\mu=1}^n
\int_0^{\l_\mu}\sqrt{{K^2(\l)-a(\l)P(\l)\over a^2(\l)}}, \eq..
$$
and the linearizing variables and flow are again defined by eqs. \Abinta,
\Lin.

   An example of such a constrained integrable system on $T^*S^{n-1}$ is
generated by the invariant
$$
\eqalignno{
\phi_R &:= -2P_{n-2} \cr
&={1\over 2}\left(\sum_{i=1}^n x_i^2\right)\left(\sum_{j=1}^n y_j^2\right)
-{1\over 2}\left(\sum_{i=1}^nx_i y_i\right)^2
 -{1\over 2}\left(\sum_{i=1}^n x_i^2\right)\left(\sum_{j=1}^n
{\mu_j^2\over x_j^2}\right) \cr
& \qquad -2 a \sum_{i=1}^n\a_i x_i y_i +
b \sum_{i=1}^n \a_i x_i^2, \eqn..}
$$
which, on the constrained manifold defined by \Constra, \Constrb becomes
$$
\phi_R = {1\over2} \sum_{i=1}^ny_i^2 -{1\over 2}{\mu_j^2\over x_j^2}
-2 a \sum_{i=1}^n \a_i x_i y_i + b \sum_{i=1}^n \a_i x_i^2.  \eq..
$$
For $a=0$, this gives the Rosochatius system \cite{Mo, GHHW, AHP}. If all
the $\mu_i$'s also vanish, it reduces to the Neumann oscillator system
\cite{Mo, H}.
\medskip
\Subtitle {1c. Case (ii) \ $c \neq 0$. Restiction to $T^*{\cal E}^{n-1}$}

 In this case, the spectral Darboux coordinates $\{\l_\mu,\z_\mu\}_{\mu=1,
\dots n}$ are defined by the relations
$$
\eqalignno{
\sum_{i=1}^n {x_i^2\over \l -\a_i} +2c &=2c {Q(\l)\over a(\l)}
\eqn Darbiia.a.\cr
Q(\l)&:= \prod_{\mu=1}^n (\l-\l_\mu) \eqn  Darbiib.b.\cr
\z_\mu&:={1\over 2}\sum_{i=1}^n {x_iy_i \over \l_\mu -\a_i}. \eqn Darbiic.c.}
$$
 The Jacobian matrix is thus again given by
$$
{\di x_i \over \di \l_\mu} ={1\over 2} {x_i\over \l_\mu -\a_i}
\eq Jacobmatrixii..
$$
and its inverse by
$$
{\di \l_\mu \over \di x_i} = -{1\over c} {a(\l_\mu)\over Q'(\l_\mu) }{x_i
\over \l_\mu -\a_i}.  \eq InvJacmatrixii..
$$
The coordinate frame fields are therefore
$$
{\di \over \di \l_\mu} ={1\over 2} \sum_{i=1}^n {x_i\over \l_\mu
-\a_i}{\di\over \di x_i}, \quad \mu=1, \dots n  \eq Framefieldii..
$$
and the canonical $1$--form is
$$
\th = \sum_{i=1}^n y_idx_i = \sum_{\mu=1}^{n}\z_\mu d\l_\mu.
\eq Canformii..
$$
To identify $\{\l_\mu\}_{\mu=1, \dots n}$ as hyperellipsoidal coordinates,
the constants $\{\a_i\}_{i=1, \dots n}$  must again be chosen as real and
positive, and an ordering fixed, e.g., by
$$
\l_n < \a_n < \l_{n-1} < \a_{n-1} < \dots < \l_1<\a_1.  \eq..
$$

The ${\bf R}^n$ euclidean metric in these coordinates is
$$
\sum_{i=1}^n dx_i^2 = -{c\over 2} \sum_{\mu=1}^{n}{Q^\prime(\l_\mu)\over
a(\l_\mu)} d\l_\mu^2 \eq Metricii..
$$
and the volume form is
$$
dV = \left({c\over 2}\right)^{n\over 2}{\prod_{\nu < \mu}^{n}(\l_\mu -
\l_\nu)\over
   [\prod_{\mu} a(\l_\mu)]^{1\over 2}}\, d\l_1 \wedge \dots \wedge d\l_{n}.
\eq Volformb..
$$
For reference in Sec. 2, we note that the scalar Laplacian is
$$
\Delta = {2\over c}\sum_{\mu=1}^{n}{a(\l_\mu)\over
Q'(\l_\mu)} \left({\di^2\over \di \l_\mu^2}
 + {1\over 2}\sum_{j=1}^n{1\over \l_\mu - \a_j}{\di\over \di\l_\mu} \right).
\eq Laplacian2..
$$

The values $\{P(\l_\mu)\}_{\mu=1, \dots n}$ are again given by
$$
{P(\l_\mu) \over a(\l_\mu)}
= - (\z_\mu -a)^2 + {K(\l_\mu)^2 \over a(\l_\mu)^2} \eq Poveraii..
$$
and, using Lagrange interpolation, $P(\l)$ may be expressed as
$$
P(\l)= -\sum_{\mu=1}^{n}{Q(\l)a(\l_\mu)\over (\l -\l_\mu)Q'(\l_\mu)}
\left( (\z_\mu -a)^2-{K(\l_\mu)^2\over a(\l_\mu)^2}\right)
- Q(\l)\left(a^2 + bc\right). \eq Lagrintii..
$$

   To obtain the Hamilton-Jacobi equation, we again reinterpret the
coefficients of the invariant polynomial $P(\l)$ in \Lagrintii as integration
constants and replace the canonical momentum components $\z_\mu $ by the
partial derivatives  ${\di S \over\di \l_\mu}$, giving
$$
-\sum_{\mu=1}^{n}{Q(\l) a(\l_\mu)\over (\l -\l_\mu)Q'(\l_\mu)}
\left(\left({\di S \over \di \l_\mu} -a\right)^2
-{K^2(\l_\mu)\over a^2(\l_\mu)}\right)
- Q(\l)\left(a^2 + bc\right) = P(\l),  \eq HamJacii..
$$
where the leading term of $P(\l)$ is $-(a^2 +bc) \l^{n}$ and the remaining
terms are independent integration constants.

  The integration of \HamJacii again proceeds by separation of variables.
Expressing $S$ in the separated form
$$
S(\l_1, \dots \l_n) = \sum_{\mu=1}^{n} s_\mu(\l_\mu), \eq Sl..
$$
dividing both sides of \HamJacii by $Q(\l)$ and equating the residues at
$\{\l=\l_\mu\}_{\mu=1, \dots n}$ gives
$$
\left({\di s_\mu \over \di \l_\mu} -a\right)^2 = {K(\l_\mu)^2 - a(\l_\mu)
P(\l_\mu) \over a(\l_\mu)^2}. \eq Separii..
$$
This may be integrated to give
$$
S(\l_1, \dots, \l_{n-1})= a\sum_{\mu=1}^{n}\l_\mu + \sum_{\mu=1}^n
\int_0^{\l_\mu}\sqrt{{K^2(\l)-a(\l)P(\l)\over a^2(\l)}}d\l.\eq..
$$
The linearizing coordinates are thus
$$
Q_i := {\di S \over \di P_i} = {1\over 2}\sum_{\mu=1}^{n} \int_0^{\l_\mu}
{\l^i \over \sqrt{ K^2(\l) - a(\l)P(\l)}}d\l, \qquad i=0, \dots n-1.
\eq Abintii..
 $$
The first $n-1$ abelian integrals in \Abintii, with $i=0, \dots n-2$, are all
of the first kind,  again defining the Abel map to the Jacobi variety
$\JJ(\CC)$, while the remaining one, giving $Q_{n-1}$, is again singular,
the integrand having a pair of simple poles over $\l=\infty$.  The linear flow
induced by any Hamiltonian $\phi=\phi(P_0, \dots P_{n-1})$ in the ring of
spectral invariants $\II^Y_{AKS}$ is, as before, given by
$$ Q_i = Q_{i0} + {\di \phi
\over \di P_i}t, \qquad i=0, \dots n-1.  \eq Linii..
$$

   We may also impose the constraints
$$
\eqalignno{
\sum_{i=1}^n {x_i^2\over \a_i} &= 2c  \eqn Constraintiia.a. \cr
\sum_{i=1}^n{x_iy_i\over \a_i} &=0     \eqn Constraintiib.b.}
$$
\nextnumber
defining the cotangent bundle $T^*\EE^{n-1}$ to the ellipsoid
$\EE^{n-1}\ss{\bf R}^n$ defined by \Constraintiia. (Here, the constant $c$
must be chosen as positive; more generally, arbitrary signs may be allowed
for the $\a_i$'s and $c$, thereby defining various hyperboloids.) In terms of
the canonical coordinates $\{\l_\mu, \z_\mu\}_{\mu=1, \dots n}$, the two
constraints are equivalent to $$
\eqalignno{
 \l_n &=0   \eqn Constraintlna.a.  \cr
\z_n &= 0. \eqn Constraintznb.b.}
$$

  For reference in Sec. 2, we note that the induced metric on
${\cal E}^{n-1}$ is
$$
\sum_{i=1}^ndx_i^2 \vert_{{\cal E}^{n-1}}=
-{s\over 2} \sum_{\mu=1}^{n-1}{\l_\mu Q^\prime(\l_\mu)\over a(\l_\mu)}
d\l_\mu^2, \eq MetricE..
$$
the corresponding volume form is
$$
dV_{{\cal E}^{n-1}} = \left({c\over 2}\right)^{n-1\over 2}
\prod_{\nu < \mu}^{n-1}(\l_\mu -\l_\nu)
\prod_{\mu=1}^{n-1}{\l_\mu^{1\over2}d\l_\mu\over|a(\l_\mu)|^{1\over 2}}
\,d\l_1 \wedge \dots \wedge d\l_{n-1}. \eq VolformE..
$$
and the scalar Laplacian is
$$
\Delta_{{\cal E}^{n-1}}= {2\over c}\sum_{\mu=1}^{n-1}{a(\l_\mu)\over
\l_\mu Q'_0(\l_\mu)} \left({\di^2\over \di \l_\mu^2}
 + {1\over 2}\left[\sum_{j=1}^n{1\over \l_\mu - \a_j}-{1\over\l_\mu}\right]
{\di\over \di\l_\mu} \right). \eq LaplacianE..
$$

Eqs. \Constraintlna, \Constraintznb determine the cotangent bundle
$T^*\EE^{n-1}\ss {\bf R}^{2n}$ as a symplectic submanifold; i.e.,  they are
purely second class, and the restriction of the remaining coordinates
$\{\l_\mu,
\z_\mu\}_{\mu=1, \dots n-1}$ provide canonical coordinates on this constrained
manifold. The restriction of the canonical $1$--form defines the canonical
$1$--form on $T^*\EE^{n-1}$
$$
\th|_{T^*\EE^{n-1}}=\sum_{\mu=1}^{n-1} \z_\mu d\l_\mu. \eq Canonellipsoid..
$$
Although neither of the constraints \Constraintiia, \Constraintiib is
individually invariant under the AKS flows generated by the invariants $P_0,
\dots P_{n-1}$, they are equivalent to the pair
$$
\eqalignno{
Q(0)&=0 \eqn QConstraint.a. \cr
{P(0)\over a(0)}={P_0\over a(0)}&= -a^2
 +{1\over 4} \left(\sum_{i=1}^n {\m_i\over \a_i}\right)^2
\eqn PConstraint.b.}
$$
and the second of these is invariant. It follows that the restrictions of the
remaining invariants $\{P_1, \dots P_{n-1}\}$ to the constrained manifold
$T^*\EE^{n-1}\ss {\bf R}^{2n}$ also Poisson commute, generating completely
integrable systems. On $T^*\EE^{n-1}$, we may write
$$
Q(\l)=\l Q_0(\l) \eq..
$$
where
$$
Q_0(\l) :=\prod_{\mu=1}^{n-1}(\l-\l_\mu). \eq..
$$

  Proceeding again by Lagrange interpolation, we have
$$
\eqalignno{
P(\l) =&\sum_{\mu=1}^{n-1}{\l a(\l_\mu) \over (\l -\l_\mu) \l_\mu
Q_0'(\l_\mu)} \left(-(\z_\mu -a)^2 +{K^2(\l_\mu) \over a^2(\l_\mu)}\right)
\cr &\quad + Q_0(\l)\left(-(a^2+bc)\l + {a(0)\over Q_0(0)}
\left(-a^2  +{1\over 4}\left(\sum_{i=1}^{n}{\mu_i^2\over
\a_i}\right)^2\right)\right), \eqn..}
$$
which, with suitable reinterpretation of $P(\l)$ in terms of integration
constants, gives the Hamilton-Jacobi equation on $T^*\EE^{n-1}$ as
$$
\eqalignno{
&\sum_{\mu=1}^{n-1}{\l a(\l_\mu) \over (\l-\l_\mu) \l_\mu Q_0'(\l_\mu)}
\left(-\left({\di S\over \di \l_\mu} -a\right)^2 + {K^2(\l_\mu) \over
a^2(\l_\mu)}\right)    \cr
&\quad+Q_0(\l)\left(-(a^2+bc)\l + {a(0)\over
Q_0(0)}  \left(-a^2  +{1\over 4}\left(\sum_{i=1}^{n}{\mu_i\over
\a_i}\right)^2\right)\right) =P(\l),  \eqn..}
$$
with $P(0)$ fixed to satisfy \PConstraint. Expressing $S(\l_1, \dots
\l_{n-1})$ in separated form as
$$
S(\l_1, \dots \l_{n-1})=\sum_{\mu=1}^{n-1} s_\mu(\l_\mu), \eq SCl..
$$
and proceeding as in the unconstrained case again gives  \Separii, for
$\mu=1,\dots n-1$, which upon integration gives
$$
S(\l_1, \dots, \l_{n-1})= a\sum_{\mu=1}^{n-1}\l_\mu + \sum_{\mu=1}^{n-1}
\int_0^{\l_\mu}\sqrt{{K^2(\l)-a(\l)P(\l)\over a^2(\l)}}d\l.\eq..
$$
The linearizing coordinates are therefore
$$
Q_i := {\di S \over \di P_i} = {1\over 2}\sum_{\mu=1}^{n-1} \int_0^{\l_\mu}
{\l^i \over \sqrt{ K^2(\l) - a(\l)P(\l)}}d\l, \qquad i=1, \dots n-1.
\eq AbintConstrainedii..
 $$
and the linear flow induced by any Hamiltonian $\phi=\phi(P_1, \dots P_{n-1})$
in the ring $\II^Y_{AKS}$ of spectral invariants is again given by
$$
Q_i = Q_{i0} + {\di \phi \over \di P_i}t, \qquad i=1, \dots n-1.  \eq
LinConstrainedii..
$$
In general, the abelian integrals in \AbintConstrainedii are all of the first
kind, so the linearizing map is again the Abel map to the Jacobi
variety $\JJ(\CC)$.

   As an example of such a constrained system within the commuting family of
systems generated by $\II^Y_{AKS}$, consider the Hamiltonian
$$
\eqalignno{
\phi_J &:= 2P_{n-1} = \sum_{i=1}^{n} I_i + (a^2 + bc)\sum_{i=1}^n\a_i \cr
   &= c\sum_{i=1}^n\left(y_i^2 -{\mu_i^2 \over x_i^2}\right)
+2a\sum_{i=1}^nx_iy_i - b\sum_{i=1}^n x_i^2. \eqn Jacellips..}
$$
Taking $c={1\over 2}$, $a=b=0$,  $\{\mu_i=0\}_{i=1, \dots n}$, this reduces
to the Jacobi system, determining geodesic motion on an ellipsoid \cite{Mo}.
More generally, if $b\neq0$, $\mu_i\neq 0$, we have motion on the
ellipsoid in the presence of both a harmonic force, directed towards the
origin, and forces  derived from the $\sum_{i=1}^n{\mu_i^2 \over x_i^2}$
potential, as in the Rosochatius system.

\section 2. Quantum Integrable Systems in $\wt{\gr gl}(2)^{+*}$
\medskip
\Subtitle {2a. Ambient Space}

  We quantize the systems considered above first in the ambient space
${\bf R}^n$, using canonical quantization within the Schr\"odinger
representation. The wave  functions will thus be taken as smooth functions
$\Psi(x_1, \dots x_n)$ on ${\bf R}^n$, square integrable with respect to the
standard measure. Using the canonical quantization rule (with
$\hbar\equiv 1$)
$$
x_j\lra \hat{x}:=x_j \qquad y_j\lra \hat{y}_j:= -i {\di \over \di x_j}
:=-i\di_j,  \eq..
$$
we obtain a quantum analogue of the loop algebra element $\NN(\l)$
$$
\NN(\l)\lra \hat{\NN}(\l):=
{1\over 4}\pmatrix{\hat{h}(\l) & \hat{e}(\l)\cr
                    \hat{f}(\l) & \hat{g}(\l)},  \eq Nhat..
$$
where
$$
\eqalignno{
\hat{h}(\l)&:= \sum_{j=1}^n{x_j\di_j + \mu_j \over \l -\a_j} +2a
 \eqn Hath.a. \cr
\hat{e}(\l)&:= \sum_{j=1}^n{\di_j^2 + {\mu_j^2\over x_j^2}\over \l-\a_j} +2b
 \eqn Hate.b.\cr
\hat{f}(\l)&:= \sum_{j=1}^n{x_j^2  \over \l -\a_j} +2c  \eqn Hatf.c.  \cr
\hat{g}(\l)&:= \sum_{j=1}^n{-x_j\di_j + \mu_j \over \l -\a_j} - 2a.
\eqn Hatg.d. }
$$
Note that we have not chosen to order these operators so as to assure
self-adjointness; they will only enter at an intermediate stage in
subsequent calculations and need not themselves be viewed as quantized
dynamic variables.  We also define an operator that plays essentially the
r\^ole of $\det{\, \NN}(\l)$ in the preceding calculations:
$$
\hat{\DD}(\l) := \hat{h}(\l) \hat{g}(\l) -\hat{f}(\l) \hat{e}(\l).  \eq
Qdet..
$$
Again, we are not concerned with whether this is an appropriately ordered
``quantum determinant'', since it will only appear as a calculational
convenience in what follows; the choice of ordering in \Qdet is the
most convenient in the subsequent Lagrange interpolation.

 A direct computation shows that $\hat{\DD}(\l)$ may be expressed as
$$
 \hat{\DD}(\l) = {1\over 2}
\sum_{i=1}^n{\hat{I}_i\over \l-\a_i}
-{1\over 4}\sum_{i=1}^n\sum_{j=1}^n{x_i\di_i\over (\l-\a_i)(\l-\a_j)}
+{1\over 2} \sum_{i=1}^n{x_i\di_i\over (\l-\a_i)^2}
+{ia\over 2} \sum_{i=1}^n{1\over \l-\a_i} -(a^2+b c)  \eq DDl..
$$
where
$$
\hat{I}_i :={1\over 2}\sum_{j=1 \atop j\neq i}^n {-(x_i\di_j - x_j \di_i)^2
-\mu_i^2{x_j^2\over x_i^2} -\mu_j^2{x_i^2\over x_j^2}
+ 2\mu_i \mu_j\over \a_i -\a_j} -2ia(x_i\di_i +{1\over2})
-b x_i^2 -c\left(\di_i^2 +{\mu_i^2\over x_i^2}\right).
  \eq..
$$
Here the $\hat{I}_i$'s {\it are} self-adjoint, and represent the quantized
version of the Poisson commuting invariants $I_i$ of the preceding section.
Moreover, a direct computation shows that these $\hat{I}_i$'s also commute
amongst themselves
$$
[\hat{I}_i, \hat{I}_j] =0 \qquad \forall \quad i,j =1, \dots n.  \eq..
$$

   Using either definition  \Darbia -- \Darbib or \Darbiia--\Darbiib for the
coordinates $\l_\mu$, and the fact that $\hat{f}(\l_\mu)=0$, it follows from
eq. \Framefieldia or \Framefieldii  that,
$$
\hat{\DD}(\l_\mu) = {\di^2 \over \di \l_\mu^2} -2ia {\di\over \di \l_\mu} +
{1\over 2}\sum_{j=1}^n {x_j\di_j \over (\l_\mu -\a_j)^2} +
{K^2(\l_\mu)\over  a^2(\l_\mu)} -a^2,   \eq  DDlmu..
$$
where it is understood that the factor $(\l_\mu -\a_j)^2$ in  the
denominator precedes the derivation $x_j\di_j$. Define the differential
operator valued polynomial $\hat{P}(\l)$ by
 $$
{\hat{P}(\l) \over a(\l)} := {1\over 2} \sum_{i=1}^n
{\hat{I}_i\over\l-\a_i}.  \eq Phat..
$$
Equating the expressions for $\hat{\DD}(\l_\mu)$ given by eqs. \DDl and
\DDlmu, gives
$$
{\hat{P}(\l_\mu)\over a(\l_\mu)}=
\left({\di \over \di\l_\mu}-ia\right)^2
+{1\over 2}\sum_{j=1}^n {1\over \l_\mu -\a_j}
\left({\di \over \di \l_\mu} -ia\right)  +{K^2(\l_\mu) \over a^2(\l_\mu)}
+a^2 +bc,  \eq Plmu..
$$
where again, the ordering on the LHS of \Plmu is understood to place the
$a(\l_\mu)$ term in the denominator {\it before} the differential operator
in the numerator. Note also, that in the definition \Phat of $\hat{P}(\l)$,
the leading term is
$$
\hat{P}(\l) = {1\over 2}\sum_{i=1}^n
\hat{I}_i\l^{n-1} + \OO(\l_{n-2}), \eq Phatasympt..
$$
where
$$
\hat{P}_{n-1}:= {1\over 2}\sum_{i=1}^n \hat{I}_i
= -{1\over 2} \sum_{i=1}^n \left( 2ia (x_i\di_i + {1\over 2}) +b x_i^2
+c(\di_i^2 +{\mu_i^2\over x_i^2})\right).  \eq Pnminusone..
$$
Thus, contrary to the definition of $P(\l)$ in \DetN, $\hat{P}(\l)$ here
does not contain the constant leading term  $-(a^2 +bc)\l^n$, and
is an operator-valued polynomial of degree $n-1$.

   In the following sections, we again must distinguish between the cases
$c=0$ and $c\neq 0$.
\medskip
\Subtitle {2b. Case (i) \ $c=0$. Restiction to $S^{n-1}$}

   Taking \Darbia --\Darbic as the definition of our coordinates
$\{\l_\mu, q\}_{\mu=1,\dots, n-1}$ and of $Q(\l)$, and using eqs. \Phat,
\Phatasympt, \Pnminusone, Lagrange interpolation gives
$$
\eqalignno{
\hat{P}(\l)= &\sum_{\mu=1}^{n-1}{Q(\l)a(\l_\mu)\over (\l-\l_\mu)Q'(\l_\mu)}
\left( \left({\di\over \di \l_\mu} -ia\right)^2
 + {1\over 2}\sum_{j=1}^n{1\over \l_\mu - \a_j} \left( {\di\over \di
\l_\mu} -ia\right) + {K^2(\l_\mu)\over a^2(\l_\mu)}+a^2\right)\cr
& -\quad Q(\l) \left( 2ia {\di \over \di q} +{n i a \over 2}
+{b\over 2}e^q \right) \eqn QLagrinti..}
$$
as the quantum analogue of \Lagrinti. Here we have used
$$
{1\over 2} \sum_{j=1}^n \hat{I}_j = -2ia {\di \over \di q} -{n i a \over 2}
-{b\over 2} e^q.  \eq QSumIji..
$$

Letting
$$
{P^E(\l)\over a(\l)}:= {1\over 2}\sum_{i=1}^n{I_i^E\over \l -\a_i}, \eq PE..
$$
where $\{I_i^E\}_{i=1,\dots n}$ is a set of simultaneous eigenvalues of the
operators $\{\hat{I}_i\}_{i=1,\dots n}$, the time independent Schr\"odinger
equation corresponding to the $1$--parameter linear family of commuting
Hamiltonians $\hat{P}(\l)$ is thus
$$
\eqalignno{
&\sum_{\mu=1}^{n-1}{Q(\l)a(\l_\mu)\over (\l-\l_\mu)Q'(\l_\mu)}
\left( \left({\di\over \di \l_\mu} -ia\right)^2
 + {1\over 2}\sum_{j=1}^n{1\over \l_\mu - \a_j} \left( {\di\over \di
\l_\mu} -ia\right) + {K^2(\l_\mu)\over a^2(\l_\mu)}+a^2\right)\Psi\cr
& \quad +Q(\l) \left( -2ia {\di \over \di q} -{n i a \over 2}
-{b\over 2}\sum_{j=1}^n x_j^2\right)\Psi
=P^E(\l)\Psi, \eqn Schrodi..}
$$
where $\Psi=\Psi(\l_1, \dots \l_{n-1},q)$.

  Choosing $\Psi$ in the completely separated form,
$$
\Psi(\l_1, \dots \l_{n-1},q)=\phi(q)\prod_{\mu=1}^{n-1}\psi_\mu(\l_\mu)
e^{ia\sum_{\mu=1}^{n-1}\l_\mu}, \eq Psisepari..
$$
dividing \Schrodi by $Q(\l)$ and equating the leading terms in $\l$
and the residues at $\l=\l_\mu$, we find that
each $\psi_\mu(\l_\mu)$ satisfies the same separated equation
$$
\left( {\di^2\over \di \l^2} +{1\over 2}\sum_{j=1}^n{1\over
\l-\a_j}{\di\over \di \l} + {K^2(\l)\over a^2(\l)}
-{1\over 2}\sum_{j=1}^n{I_j^E\over \l-\a_j} +a^2\right)\psi(\l) = 0,
\eq Schrodsepa.a.
$$
while $\phi(q)$ satisfies
$$
\left(-2ia{\di\over \di q} -{nia\over 2} -{b\over 2} e^q\right)\phi(q)
={1\over 2}\sum_{j=1}^nI_j^E \phi(q). \eq Schrodsepb.b.
$$
Eq. \Schrodsepa is a Fuchsian differential equation of the generalized Lam\'e
type already encountered in previously studied examples of such systems
\cite{BT, Mc}. Note also that, if $\l$ and $\z$ are viewed as canonically
conjugate variables, \Schrodsepa may be interpreted as the quantized form of
the
characteristic equation \Spectralcurve  defining the classical spectral curve
$\CC$. Thus, the completely separated form of the Schr\"odinger equation may be
viewed as the ``quantized spectral curve'' (cf. \cite{Sk3}).

   If $a=0$, $q$ is a conserved quantity and if $b\neq 0$, eq. \Schrodsepb
should be interpreted as fixing the constant value of $q$ in terms of the
invariant $\sum_{i=1}^nI_j^E$. Then eq. \Schrodsepb implies that,
rather than choosing $\phi$ to be an $L^2$ function, it should  be chosen
as a delta function within the coordinate representation in which $q$ is
fixed by the eigenvalues $I_j^E$. If $a=0$ and $b=0$, this choice
may still be made, but the eigenvalues $I_j^E$ are not all independent,
since they must sum to $0$,
$$
\sum_{i=1}^n I^E_j \equiv 0,  \eq..
$$
but the invariant $q$ may be set equal to any constant value independently.

   Whether $a=0$ or not, we may impose the second class constraints
\Constrq, \Constrp in the quantum problem by requiring the wave function to
satisfy
$$
\left( {\di \over \di q} + {n\over 4}\right) \Psi \vert_{q=0}=0,
\eq QConstri..
$$
which is the quantum analogue of the constraint \Constrp, and
setting $q=0$ in the expression \QLagrinti defining $\hat{P}(\l)$.
Restricting to the subspace consisting of $\Psi$'s  satisfying \QConstri,
the coefficients of $\hat{P}(\l)$ generate an $n-1$ dimension linear space
of commuting operators on the sphere $S^{n-1}$. This is spanned, e.g., by
the restrictions of the operators $\hat{I}_j$, which, because of \QConstri
satisfy
$$
\sum_{j=1}^n\hat{I}_j = -b.  \eq..
$$

Applying the constraint, \Schrodi reduces to
$$
\eqalignno{
&\sum_{\mu=1}^n{Q(\l)a(\l_\mu)\over (\l-\l_\mu)Q'(\l_\mu)}
\left( \left({\di\over \di \l_\mu} -ia\right)^2
 + {1\over 2}\sum_{j=1}^n{1\over \l_\mu - \a_j} \left( {\di\over \di
\l_\mu} -ia\right) + {K^2(\l_\mu)\over a^2(\l_\mu)}+a^2\right)\Psi_0 \cr
&\quad -b Q(\l)\Psi_0
=P^E(\l)\Psi_0, \eqn ConstrSchrodi..}
$$
where
$$
\Psi(\l_1, \dots \l_{n-1}, q) := e^{-{nq\over 4}} \Psi_0(\l_1,\dots
\l_{n-1})  \eq..
$$
and the eigenvalues $I_j^E$ must satisfy
$$
\sum_{j=1}^nI_j^E=-b.  \eq..
$$
The volume form on $S^{n-1}$ in these coordinates is given by \VolformS
and the polynomial family of operators appearing on the LHS of
\ConstrSchrodi are all self-adjoint with respect to the corresponding
scalar product. The separation of variables for the resulting systems on
the constrained space $S^{n-1}$ is again obtained by setting
$$
\Psi_0(\l_1,\dots \l_{n-1}) =\prod_{i=1}^{n-1}\psi_\mu(\l_\mu)
e^{ia\sum_{\mu=1}^{n-1}\l_\mu}
\eq PsiConstrsepari..
$$
in \ConstrSchrodi, dividing by $Q(\l)$ and equating residues at
$\l=\l_\mu$. The resulting separated equations for the $\psi_\mu(\l_\mu)$
are all again given by \Schrodsepa. The particular case of this family of
systems corresponding to choosing all $\mu_i=0$ and $a=0$, as in the
classical case, gives the quantum Neumann oscillator system (cf. \cite{BT}),
while if the $\mu_i$'s are non-zero, we obtain the quantum Rosochatius system
(cf. \cite{Mc}).
\medskip
\Subtitle {2c. Case (ii) \ $c \neq 0$. Restiction to ${\cal E}^{n-1}$}

   We begin again with the unconstrained system in the ambient space
$\bf{R}^n$. Taking \Darbiia, \Darbiib as the definition of the coordinates
$\{\l_\mu\}_{\mu=1, \dots n}$, and of $Q(\l)$,  and again using
eqs. \Phat--\Pnminusone, Lagrange interpolation gives
$$
\eqalignno{
&\hat{P}(\l)= \eqn QLagrintii..\cr
 &\quad \sum_{\mu=1}^{n}{Q(\l)a(\l_\mu)\over
(\l-\l_\mu)Q'(\l_\mu)}
 \left( \left({\di\over \di \l_\mu} -ia\right)^2
 + {1\over 2}\sum_{j=1}^n{1\over \l_\mu - \a_j} \left( {\di\over \di
\l_\mu} -ia\right) + {K^2(\l_\mu)\over a^2(\l_\mu)}+a^2+bc \right)}
$$
as the quantum analogue of \Lagrintii.
Defining the polynomial family of eigenvalues $P^E(\l)$ again by eq. \PE
in terms of the set $\{I_i^E\}_{i=1,\dots n}$of simultaneous eigenvalues of the
operators $\{\hat{I}_i\}_{i=1,\dots n}$, the time independent Schr\"odinger
equation corresponding to the $1$--parameter linear family of commuting
Hamiltonians $\hat{P}(\l)$ is now
$$
\eqalignno{
&\sum_{\mu=1}^n{Q(\l)a(\l_\mu)\over (\l-\l_\mu)Q'(\l_\mu)}
 \left( \left({\di\over \di \l_\mu} -ia\right)^2
 + {1\over 2}\sum_{j=1}^n{1\over \l_\mu - \a_j} \left( {\di\over \di
\l_\mu} -ia\right) + {K^2(\l_\mu)\over a^2(\l_\mu)}+a^2 + bc\right)\Psi\cr
&\quad=P^E(\l)\Psi, \eqn Schrodii..}
$$
where $\Psi=\Psi(\l_1, \dots \l_{n})$.
 Choosing $\Psi$ in the completely separated form,
$$
\Psi(\l_1, \dots \l_{n})= e^{ia\sum_{\mu=1}^{n}\l_\mu}
\prod_{\mu=1}^{n}\psi_\mu(\l_\mu), \eq Psiseparii..
$$
dividing \Schrodi by $Q(\l)$ and equating the residues at $\l=\l_\mu$, we
find again that each $\psi_\mu(\l_\mu)$ satisfies the same separated
equation
$$
\left( {\di^2\over \di \l^2} +{1\over 2}\sum_{j=1}^n{1\over
\l-\a_j}{\di\over \di \l} + {K^2(\l)\over a^2(\l)}
-{1\over 2}\sum_{j=1}^n{I_j^E\over \l-\a_j} +a^2 + bc\right)\psi(\l) = 0.
\eq Schrodsepii..
$$

In the the quantum problem corresponding to the system with second
class constraints \Constraintlna \Constraintznb, it is
preferable to impose the invariant constraint (cf. eq. \PConstraint)
$$
\hat{P}(0)\Psi =P^E(0)\Psi,  \eq QConstraintE..
$$
where
$$
{P^E(0)\over a(0)} = -{1\over2}\sum_{i=1}^n{I_i^E\over \a_i} =
{K^2(0)\over a^2(0)}+bc. \eq QConstrainteigenvalE..
$$
Letting
$$
\Delta_\mu:= \left({\di\over\di \l_\mu}-ia\right)^2
+{1\over 2} \sum_{j=1}^n {1\over \l_\mu-\a_j}
\left({\di\over \di \l_\mu} -ia\right) + {K^2(\l_\mu)\over a^2(\l_\mu)}
+a^2 +bc,  \eq Deltamu..
$$
we have
$$
\eqalignno{
\hat{P}(\l) &={Q_0(\l) a(\l_n) \over Q_0(\l_n)}\Delta_n
+ \sum_{\mu=1}^{n-1} {(\l-\l_n)Q_0(\l)a(\l_\mu)\over
    (\l-\l_\mu)(\l_\mu - \l_n) Q'_0(\l_\mu)} \Delta_\mu \cr
&= Q_0(\l)\left[ \hat{R}(\l)-\hat{R}(\l_n) + {a(\l_n)\over
Q_0(\l_n)} \Delta_n\right],
 \eqn Phatc..}
$$
where
$$
\hat{R}(\l) := \sum_{\mu=1}^{n-1} {a(\l_\mu) \over (\l-\l_\mu)
Q'_0(\l_\mu)} \Delta_\mu.  \eq Rhatc..
$$
It follows that, for $\Psi$ satisfying the constraint \QConstraintE,
we have
$$
\hat{P}(\l)\Psi = \hat{P}_0 (\l)\Psi,  \eq PhatredE..
$$
where
$$
\eqalignno{
\hat{P}_0(\l) &:= Q_0(\l)\left(\hat{R}(\l) - \hat{R}(0)
 + {P^E(0)\over Q_0(0)} \right) \cr
&=Q_0(\l) \left[ \sum_{\mu=1}^{n-1} {\l a(\l_\mu) \over (\l-\l_\mu)\l_\mu
Q'_0(\l_\mu)}\Delta_\mu + {P^E(0)\over Q_0(0)}\right]. \eqn PohatE..}
$$
Since the coordinate $\l_n$ does not enter in eq. \PhatredE, it may be
treated as a parameter, and set equal to $0$ in $\Psi$. This may be viewed as
the gauge
condition associated to the first class constraint \QConstraintE. Since the
constraint
commutes with the family of operators $\hat{P}(\l)$, the reduced operators
$\hat{P}_0(\l)$ defined by \PohatE still commute amongst themselves. The wave
function
$\Psi$, with $\l_n=0$, is defined on the ellipsoid  ${\cal E}^{n-1}$, and
satisfies the
polynomial family of Schr\"odinger equations
$$
Q_0(\l) \left[ \sum_{\mu=1}^{n-1} {\l a(\l_\mu) \over (\l-\l_\mu)\l_\mu
Q'_0(\l_\mu)}\Delta_\mu + {P^E(0)\over Q_0(0)}\right]\Psi=P^E(\l)\Psi,
\eq SchrodEa..
$$
where $P^E(0)$ is given by \QConstrainteigenvalE. The coefficients of
$\hat{P}_0(\l)$ thus generate an $n-1$-parameter family of commuting
invariants, which may be simultaneously diagonalized. The
simultaneous eigenfunction $\Psi$ is independent of the parameter $\l$, so
dividing \SchrodEa by $Q_0(\l)$ and taking the limit $\l\ra \infty$ gives
$$
\hat{R}(0)\Psi = \left( {P^E(0)\over Q_0(0)} -
{1\over 2}\sum_{i=1}^n I_i^E\right)\Psi,  \eq..
$$
and hence
$$
\hat{P}_0(\l) = Q_0(\l)\left(\hat{R}(\l) +\sum_{i=1}^nI_i^E\right)\Psi.
\eq..
$$
The Schr\"odinger equation \SchrodEa may therefore equivalently be written
$$
Q_0(\l) \left[ \sum_{\mu=1}^{n-1} { a(\l_\mu) \over (\l-\l_\mu)\l_\mu
Q'_0(\l_\mu)}\Delta_\mu +\sum_{i=1}^nI_i^E\right]\Psi= P^E(\l)\Psi.
\eq SchrodEb..
$$
Either way, $\Psi(\l_1,\dots \l_{n-1})$ may be chosen in completely
separated form
$$
\Psi(\l_1, \dots \l_{n-1})=e^{ia\sum_{\mu=1}^{n-1}\l_\mu}
\prod_{\mu=1}^{n-1}\psi_\mu(\l_\mu), \eq PsiseparE..
$$
where each $\psi_{\mu}(\l_\mu)$ still satisfies \Schrodsepii.

Notice, however, that the reduced operator $\hat{P}_0(\l)$ obtained in
this way is not self-adjoint with respect to the scalar product
$$
\langle\Phi,\ \Psi\rangle
:=\int_{{\cal E}^{n-1}}\bar{\Phi}\Psi dV_{{\cal E}^{n-1}},
 \eq Scalarprod..
$$
with $dV_{{\cal E}^{n-1}}$ the standard volume form \VolformE on
${\cal E}^{n-1}$. It is, however, with respect to the scalar product
$$
\widetilde{\langle\Phi,\ \Psi\rangle} := \int_{{\cal E}^{n-1}}\bar{\Phi}\Psi
d\wt{V}_{{\cal E}^{n-1}} \eq Scalarprodtilde..
$$
where $d\wt{V}_{{\cal E}^{n-1}}$ is the modified volume form defined by
$$
d\wt{V}_{{\cal E}^{n-1}} = \left({c\over 2}\right)^{n-1\over 2}
\prod_{\nu < \mu}^{n-1}(\l_\mu -\l_\nu)
\prod_{\mu=1}^{n-1}{\l_\mu d\l_\mu\over|a(\l_\mu)|^{1\over 2}}
\,d\l_1 \wedge \dots \wedge d\l_{n-1}. \eq VolformEE..
$$
Making the transformation
$$
\eqalignno{
\Psi &\lra \wt{\Psi} := \prod_{\mu=1}^{n-1}\l_\mu^{1\over4}\Psi\cr
\hat{P}_0&\lra \wt{P}_0:= \prod_{\mu=1}^{n-1}\l_\mu^{1\over4}\circ
\hat{P}_0 \circ \prod_{\mu=1}^{n-1}\l_\mu^{-{1\over4}}
\eqn Transf..}
$$
gives the equivalent Schr\"odinger equation
$$
\wt{P}_0(\l)\wt{\Psi}=P^E(\l)\wt{\Psi},  \eq..
$$
where the polynomial family of operators $\wt{P}_0(\l)$,
given by
$$
\eqalignno{
\wt{P}_0(\l) :=
\sum_{\mu=1}^{n-1}&{\l Q(\l)a(\l_\mu)\over
(\l-\l_\mu)\l_\mu Q'(\l_\mu)}
\Biggl[ \left({\di\over \di \l_\mu} -ia\right)^2
 + {1\over 2}\left(\sum_{j=1}^n{1\over \l_\mu - \a_j}-{1\over
\l_\mu}\right) \left( {\di\over \di \l_\mu} -ia\right)\cr
&+ {K^2(\l_\mu)\over a^2(\l_\mu)}
+a^2+bc+{\cal U}(\l_\mu)\Biggr],
\eqn Phattildeo..}
$$
with
$$
{\cal U}(\l_\mu):= {5\over 16 \l_\mu^2} -
{1\over\l_\mu}\sum_{j=1}^n{1\over\l_\mu-\a_j},
\eq Upotential..
$$
is now self-adjoint with respect to \Scalarprod.

The terms ${\cal U}(\l_\mu)$ in \Phattildeo are of order
${\cal O}(\hbar^2)$, and hence disappear in the semi-classical limit. If we
subtract these, we obtain the polynomial family of self adjoint operators
$$
\eqalignno{
\hat{P}_c(\l) :=
\sum_{\mu=1}^{n-1}&{\l Q(\l)a(\l_\mu)\over
(\l-\l_\mu)\l_\mu Q'(\l_\mu)}
\Biggl[ \left({\di\over \di \l_\mu} -ia\right)^2
 + {1\over 2}\left(\sum_{j=1}^n{1\over \l_\mu - \a_j}-{1\over
\l_\mu}\right) \left( {\di\over \di \l_\mu} -ia\right)\cr
&+ {K^2(\l_\mu)\over a^2(\l_\mu)}
+a^2+bc \Biggr],
\eqn Phattildec..}
$$
which have the same semi-classical limit as \Phattildeo. The coefficients of
$\hat{P}_c(\l)$ may again be simultaneously diagonalized by separation of
variables in the corresponding Schr\"odinger equation $$
\hat{P}_c(\l)\Psi_c=P_c^E(\l)\Psi_c, \eq SchrodEc..
$$
where the eigenvalues $I_i^{E,c}$, defined by
$$
{P^E_c(\l)\over a(\l)}:= \sum_{i=1}^n{I_i^{E,c}\over \l -\a_i}, \eq..
$$
are again constrained to satisfy
$$
{P^E_c(0)\over a(0)} = -{1\over2}\sum_{i=1}^n{I_i^{E,c}\over \a_i} =
{K^2(0)\over a^2(0)}+bc. \eq QConstrainteigenvalE..
$$
Expressing $\Psi_c$ in the completely separated form
$$
\Psi(\l_1, \dots \l_{n})=e^{ia\sum_{\mu=1}^{n-1}\l_\mu}
\prod_{\mu=1}^{n-1}\psi_{c,\mu}(\l_\mu), \eq PsiseparEc..
$$
each $\psi_{c,\mu}(\l_\mu)$ must satisfy
$$
\left( {\di^2\over \di \l^2} +{1\over 2}\left(\sum_{j=1}^n{1\over
\l-\a_j}-{1\over \l}\right){\di\over \di \l} + {K^2(\l)\over a^2(\l)}
-{1\over 2}\sum_{j=1}^n{I_j^E\over \l-\a_j} +a^2 + bc\right)\psi_c(\l) = 0.
\eq SchrodsepEc..
$$

Since the self-adjoint operators $\wt{P}_0(\l)$ and $\hat{P}_c(\l)$ have the
same
semi-classical limits, and both are completely integrable through separation of
variables, either may be regarded as a valid quantization of the corresponding
constrained classical system on  ${\cal E}^{n-1}$.  Ambiguities of order
${\cal O}(\hbar^2)$ are well known to occur in the constrained quantization
procedure \cite{D}, and can only be resolved by appealing to some further
physical principle. In the case $a=b=\mu_i=0$, the degree $n-1$ coefficient in
$\hat{P}_c(\l)$, which in the classical system generates free motion on ${\cal
E}^{n-1}$ (cf. eq. \Jacellips), gives the Laplacian on ${\cal E}^{n-1}$,
whereas
that in $\wt{P}_0(\l)$ contains additional ${\cal O}(\hbar^2)$ potential terms
due to the ${\cal U}(\l_\mu)$ factors in \Phattildeo. This may be a valid
reason
to regard $\hat{P}_c(\l)$ as the correct quantization of this family of
integrable systems.
\bigskip \bigskip
\section 3. Discussion
\medskip

  We have seen that in the ambient space ${\bf R}^n$ and on the sphere, the
quantization of the integrable systems of Section 1 proceeds straightforwardly,
whereas on the ellipsoid, constrained quantization leads to ${\cal O}(\hbar^2)$
ambiguities. The origin of the problem may be understood by noting that,
whereas
in the coordinates of Sections 1b, 2b, which are adapted to the sphere, not
only
does separation with respect to the transversal coordinate $q$ hold for the
ambient space Schr\"odinger equation, but also for the volume form \Volforma,
giving rise to the standard sphere volume form \VolformS upon factorization. In
the case of the ellipsoid, however,  although the Schr\"odinger equation again
separates in the adapted coordinates of Sections 1c, 2c, the volume form
\Volformb
does not admit such a factorization with respect to the transversal coordinate
$\l_n$. If one just imposes the constraint $\l_n=0$ defining the ellipsoid, and
eliminates the $d\l_n$ factor in \Volformb, the volume form
$d\wt{V}_{{\cal E}^{n-1}}$ of eq. \VolformEE results, not the standard one
\VolformE determined by the induced metric. More generally, when the
constrained
manifold is obtained as an orbit in configuration space under a large symmetry
group, and the commuting operators of the system are expressible, as for the
sphere, in terms of invariant operators in the enveloping algebra, the
constraining procedure may compatibly be deduced through separation of the
remaining ``radial'' variable. This is the case, for example, in the case of
free motion on symmetric spaces \cite{BKW1, BKW2, K, KM, ORW}. For the
ellipsoid,
or more general constrained systems in which no such transitive symmetry group
is
present, further considerations regarding self-adjointness invariably
enter, and ${\cal O}(\hbar^2)$ ambiguities of the type encountered here may
arise.

  The systems studied here (prior to the constraints) may, in view of the form
of the operators $\hat{\NN}(\l)$ of eq. \Nhat, and the choice of commuting
invariants given by its ``quantum determinant'', be regarded equivalently as
defining a noncompact version of the Gaudin spin lattice \cite{G, Sk1--Sk3, Ku}
(with $su(2)$ replaced by $sl(2, {\bf R})$). From this viewpoint, the
separation
of variables may be interpreted as a ``functional Bethe ansatz''
(cf. \cite{Sk1--Sk3}).

  Two types of generalization of the systems studied here naturally suggest
themselves, both at the classical and quantum levels. The first is to allow
degeneration of the parameters $\{\a_1, \dots \a_n\}$ defining the image of the
Poisson map \JAb, \NM, including the possibility of higher order poles. This is
known (cf. \cite{K, Ku}) to lead to separable coordinates other than the
generic,
hyperellipsoidal ones appearing here, adapted to symmetry groups larger than
the finite group  $({\bf Z}_2)^n$ encountered in Section 1. The second
generalization consists of extending the present considerations to higher rank
algebras, such as ${\gr{gl}}(r)$, and reductions thereof. Whereas the classical
systems so arising are known to be integrable and separable under generic
assumptions regarding the rational coadjoint orbits and initial data
\cite{AHH4},
very little is known about the corresponding quantum systems. These questions
merit further study, and will be addressed in future work.

\bigskip \bigskip  \noindent{\it Acknowledgements.} The authors have benefited
from helpful discussions with C. Duval, E. Sklyanin, E. Kalnins and M. del Olmo
concerning this work.
\bigskip \bigskip
\references

Ad& Adler, M., ``On a Trace Functional for Formal
Pseudo-Differential Operators and the Symplectic Structure of the Korteweg-
de Vries Equation'', {\it Invent. Math.} {\bf 50}, 219--248, (1979).

AHH1& Adams, M.R., Harnad, J. and Hurtubise, J.,
``Isospectral Hamiltonian Flows in Finite and Infinite Dimensions II.
Integration of Flows'',
 {\it Commun. Math. Phys.} {\bf 134}, 555--585 (1990).

AHH2& Adams, M.R., Harnad, J. and Hurtubise, J.,
``Dual Moment Maps to Loop Algebras'', {\it Lett. Math. Phys.} {\bf 20},
 294--308 (1990).

AHH3& Adams, M.R., Harnad, J. and Hurtubise, J.,
``Liouville Generating Function for Isospectral Hamiltonian
Flow in Loop Algebras'', in:  {\it Integrable and
Superintergrable Systems}, ed. B.A.  Kuperschmidt, World Scientific,
Singapore (1990).

AHH4& Adams, M.~R., Harnad,~J. and  Hurtubise,~J., ``Darboux
Coordinates and Liouville-Arnold Integration  in Loop Algebras,''
  {\it Commun. Math. Phys.} {\bf 155}, 385-413 (1993).

AHP& Adams, M.R., Harnad, J. and Previato, E., ``Isospectral
Hamiltonian Flows in Finite and Infinite Dimensions
I. Generalised Moser Systems and Moment Maps into Loop Algebras'',
{\it Commun. Math. Phys.} {\bf 117}, 451--500 (1988).

AvM& Adler, M.\ and van Moerbeke, P., ``Completely Integrable
Systems, Euclidean Lie Algebras, and Curves,''  {\it Adv. Math.}
{\bf 38}, 267--317 (1980); ``Linearization of Hamiltonian Systems, Jacobi
Varieties and Representation Theory,''
{\it ibid.} {\bf 38}, 318--379 (1980).

BKW1&  Boyer, C.P., Kalnins, E.G. and Winternitz, P. ``Completely Integrable
Relativistic Hamiltonian Systems and Separation of variables in Hermitian
Hyperbolic Spaces'', {\it J. Math. Phys.} {\bf 24}, 2022--2034 (1983).

BKW2&  Boyer, C.P., Kalnins, E.G. and Winternitz, P. ``Separation of
Variables for the Hamilton-Jacobi Equation on Complex Projective Spaces'',
{\it SIAM J. Math. Anal.} {\bf 16}, 93--109 (1985).

BT&  Babelon, O. and Talon, M. ``Separation of variables for the classical and
quantum Neumann model'', {\it Nucl. Phys.} {\bf B379}, 321 (1992).

D& Dirac, P.A.M., {\it Lectures on Quantum Mechanics}, Belfer Graduate School
of Science Monograph Seris No. 2 (New York, 1964).

Du& Dubrovin, B.A., ``Theta Functions and Nonlinear Equations'',
 {\it Russ. Math. Surv.} {\bf 36}, 11--92 (1981).

G& Gaudin, M., ``Diagonalization d'-une classe d'hamiltoniens de spin'',
 {\it J. Physique} {\bf 37}, 1087-1098 (1976).

GH& Griffiths, P.  and Harris, J.  {\it Principles of Algebraic Geometry},
Wiley, New York (1978).

GHHW& Gagnon, L., Harnad, J.,  Hurtubise, J. and Winternitz, P.,
``Abelian Integrals and the Reduction Method for an Integrable Hamiltonian
System'', {\it J. Math. Phys.} {\bf 26}, 1605--1612 (1985).

H&  Harnad,~J.  ``Isospectral Flow and Liouville-Arnold Integration in Loop
Algebras,'' in: {\it Algebraic Geometric Methods in Mathematical Physics},
Springer Lecture Notes in Physics (1993), ed. G. Helminck.

HW&  Harnad, J. and  Wisse, and M.-A. ``Isospectral Flow in Loop Algebras
   and Quasiperiodic Solutions 	to the Sine-Gordon Equation'',
  {\it J. Math. Phys.} {\bf 34},  3518--3526 (1993).

K& Kalnins, E.G. {\it Separation of Variables for Riemannian Symmetric
Spaces of Constant Curvature}, Pitman Monographs and Surveys in Pure and
Applied
Mathematics {\bf 28} (1986).

\break
KM&  Kalnins, E.G. and Miller, W. Jr., ``Separation of Variables on
$n$--dimensional \break Riemannian manifolds: 1. The $n$--sphere $S_n$ and
Euclidean Space $R_n$'', {\it J. Math. Phys.} {\bf 27}, 1721--1736 (1986).

KMW&  Kalnins, E.G., Miller, W. Jr. and Winternitz, P. ``The Group
$O(4)$, Separation of Variables and the Hydrogen Atom'',
 {\it SIAM J. Appl. Math.} {\bf 30}, 630--664 (1976).

KN& Krichever, I.M. and  Novikov, S.P.,  ``Holomorphic Bundles over
Algebraic Curves and Nonlinear Equations'', {\it Russ. Math. Surveys}
{\bf 32}, 53--79 (1980).

Ko& Kostant, B., ``The Solution to a Generalized Toda Lattice and
Representation \break Theory'', {\it Adv. Math.} {\bf 34}, 195-338 (1979).

Ku&  Kuznetsov, Vadim B., ``Equivalence of two graphical calculi'',
{\it J. Phys. A} {\bf 25}, 6005--6026 (1992).

Mc& Macfarlane, A.~J., ``The quantum Neumann model with the
potential of \hfil \break Rosochatius'', {\it Nucl. Phys.} {\bf B386}, 453-467
(1992).

Mo&  Moser, J.,  ``Geometry of Quadrics and Spectral Theory", {\it
The Chern Symposium, Berkeley, June 1979}, 147--188, Springer,
 New York, (1980).

ORW& del Olmo, M.A.,  Rodriguez, M.A. and  and Winternitz, P.
``Integrable Systems Based on $SU(p,q)$ Homogeneous Manifolds''
{\it J. Math Phys.} {\bf 34}, 5118--5139 (1993).

Sk1& Sklyanin, E.K., ``Separation of Variables in the Gaudin Model'',
{\it J. Sov. Math.} {\bf 47} 2473--2488 (1989).

Sk2& Sklyanin, E.K., ``Functional Bethe Ansatz'',  in:  {\it Integrable and
Superintergrable \break Systems}, ed. B.A. Kuperschmidt, World Scientific,
Singapore (1990).

Sk3& Sklyanin, E.K., ``Separation of Variables in the Quantum Integrable
Models Related to the Yangian ${\cal Y}[sl(3)]$'', preprint NI-92013'' (1992).

Sy& Symes, W., ``Systems of Toda Type, Inverse Spectral Problems
and Representation Theory'', {\it Invent. Math.} {\bf 59}, 13-51 (1980).

TW&Tafel, J. et Wisse, M.A., Loop Algebra Approach to Generalized
Sine-Gordon \break Equations. {\it J. Math. Phys.} {\bf 34}, (1993)

To& Toth, John A., ``Various Quantum Mechanical Aspects of Quadratic Forms'',
 M.I.T. preprint (1992).

W& Wisse, M.A., ``Darboux Coordinates and Isospectral Hamiltonian
Flows for the Massive Thirring Model'',
{\it Lett. Math. Phys.} {\bf 28} 287--294 (1993).

\endreferences

\end